\documentclass[aps,prl,twocolumn,groupedaddress,showpacs]{revtex4}

\begin{document}


\title{
Vibration of the Dimer on Ge(001) Surface  Excited Coherently
by STM Current
}



\author{Hiroshi  Kawai}
\author{Osamu Narikiyo}


\affiliation{Department of Physics, Faculty of Sciences, Kyushu University,
Ropponmatsu, Fukuoka 810-8560, Japan}


\date{\today}

\begin{abstract}
The vibration of the dimer on Ge(001) surface 
with the higher vibrational
number excited coherently by STM current is theoretically investigated.
The coherent excitation rate of the dimer vibration
is obtained by
the Hamiltonian consisting of the terms
of the electron system and the electron-vibration coupling terms.
The transformation of the local structures reported in
STM experiments is shown to be driven 
by the dimer vibration excited coherently 
by the STM current. 
The sample bias voltage of STM 
above which the transtormation to the p(2$\times$2) structure
is able to be observed in the experiments, is semiquantitatively reproduced 
by the quasi-one-dimensional character of the $\pi^*$-band.
We show that
the excitation rate has the term not  decaying 
on the distance from the
STM tip for the one-dimensional band. 
The contrastive diffrence of the shapes of the transformed region depending
on the sign of the bias voltage observed by the experiments
is explained by the difference between
the quasi-one-dimensional character of  the $\pi^*$-band and 
the two-dimensional character of 
the $\pi$-band on Ge(001) surface.
\end{abstract}


\maketitle


On Ge(001) surface, neighboring atoms form the buckled dimers.
The orientational arrangement of the buckled dimers
orders into the c(4$\times$2) structure 
through the order-disorder phase transition 
about 315K~\cite{ferrer1,lucas1,sato1,yoshimoto1,yoshimoto2,nakamura1,kawai2}.
Recently, the phase manipulation of Ge(001) surface 
on Sb-doped substrate by 
scanning tunneling microscopy (STM)
apparatus has been reported~\cite{komori1,komori2}.
The local surface structure on Ge(001) surface is modified reversibly
and hysteretically
between c(4$\times$2) and p(2$\times$2) structures at 80K by 
the sample bias voltage $V_s$ of STM.
The transformation of the local structure
from c(4$\times$2) to p(2$\times$2)
is observed at positive sample bias voltages of $0.8{\rm V} \le V_s $ and 
the reverse transformation from p(2$\times$2) to c(4$\times$2) is done at 
negative $V_s \le -0.7{\rm V}$~\cite{komori1,komori2}. 
The transformation takes place occasionally
during the scan of the surface by STM.
The transformation rate to the p(2$\times$2) structure is reported to be
proportional to the tunneling current of STM ($I_{\rm STM}$)~\cite{komori1,komori2}.
The occasional transformation to the p(2$\times$2) structure
during the scan 
is restricted  to the dimer row including the dimer beneath the
tip of STM, and takes place simultaneously in the long region 
in the dimer row. 
In most cases, the transformed  p(2$\times$2) region
in the host structure of c(4$\times$2) continues to the outside of the
scanning area along the dimer row.
The reverse transformation to c(4$\times$2)
presents a remarkable contrast to the transformation to p(2$\times$2).
The transformed c(4$\times$2) region expands over 
the consecutive dimer rows beneath the tip.
The region, however, is restricted within the range of 
a few nm away from the dimer beneath the tip
for both directions of along  and perpendicular to the dimer row.

In the present letter, we investigate theoretically the vibration of 
the dimer on Ge(001) surface  excited coherently
by STM current, using the Hamiltonian consisting of the term
of the electron system describing the tunneling current of electrons
and the electron-vibration coupling term.
The excitation rate of the vibration of the dimer
beneath the tip to the higher vibrational
number $n$ by the coherent process
is shown to be proportional 
to $I_{\rm STM}$. 
We report that the excitation rate of the vibrartion of 
the dimer away from the tip
decays much slowly on the distance from the tip
for the direction along the dimer row
in the case of $V_s > 0$. 
The range of the distance that the transformation
is effective, is expected to be of the order of 
the mean free path of electron in the $\pi^*$-band.

Previously,
we have investigated the vibration of the dimer excited by STM current
on Si(001) surface of the p-type substrate,
where the case of negative $V_{s}$ has been treated~\cite{kawai1}. 
The vibrartion of the dimer beneath the tip on Si(001) surface 
is excited  to the higher vibrational number $n$
by the STM current
through the incoherent ladder climbing of the vibrational states
at low temperatures.
The effective temperature of the vibration on the dimer  beneath the tip
strongly deviates from the substrate temperature $T$
and reaches a few hundred Kelvin for the typical values of STM current
at low temperatures lower than 20K
on the boron doped (B-doped) Si(001) substrate.
The high effective temperatures at low $T$
cause the rapid flip-flop motion of the dimer on 
Si(001) substrate.
At these low temperatures lower than 20K,
the excitation rate of dimer vibration with the higher vibrational
number $n$ depends on the  $I_{\rm STM}$ in highly nonlinear manner
on B-doped Si(001) substrate~\cite{kawai1}.
When $T$ increases around 20K, the deexcitation of the
vibration by the transition
from the vibrational number $n$ to $n-1$
through the inner-band excitation of the electron-hole
pair creation in the $\pi$-band becomes prominent 
and 
the rate of the deexcitation increase steeply in $T$~\cite{kawai1}.
The high increase of the deexcitation rate causes
the steep decrease of the probability of
the incoherent ladder climbing of the vibrational states 
and that of the effective temperature of the dimer 
beneath the tip on Si(001) substrate.
At temperatures higher than about 35 K,
the incoherent ladder climbing excited by STM current
is not dominant 
on the excitation to the higher vibrational number $n$~\cite{kawai1}.
We have shown that the temperature where
the inner-band excitation becomes prominent 
is determined by 
the absolute value of the energy of the $\pi$-band top  
measured from the Fermi level of the electron system~\cite{kawai1}.
On the B-doped Si(001) substrate, 
the absolute value of the energy is about 22.5meV. 

On Sb-doped Ge(001) substrate used by the experiment~\cite{komori1,komori2},
the absolute value of the energy of the $\pi^*$-band bottom  
measured from the Fermi level of the electron system is about 4.8meV.
From this small value of the energy, 
the rate of
the deexcitation of the
vibration from $n$ to $n-1$
through the inner-band excitation  
in the $\pi^*$-band on the Sb-doped Ge(001) substrate
is expected to become prominent 
at temperature much lower than that on the B-doped Si(001)
substrate. The rate is estimated  to be effective at temperature
of the order of (4.8/22.5)$\times$20K$\sim$4K.
From this estimation of temperature, we know that 
the incoherent ladder climbing of the vibrational states
excited by the STM current on Sb-doped Ge(001) substrate
is  suppressed  at higer temperatures, and becomes negligible
already at 8 K. 
Therefore, 
the transformation of the local structure observed by 
the experiment at 80K and 10K
on the  Sb-doped Ge(001) substrate~\cite{komori1,komori2} 
can not be ascribed
the flip-flop motion of the dimer 
through
the incoherent ladder climbing of the vibrational states
excited by the STM current.

We introduce the Hamiltonian $H=H_{\rm e} + H_{\rm ev}$ 
representing the coupling
between the electronic system and the localized vibrational
system,
\begin{eqnarray}
H_{\rm e} &=& 
\sum_k 
\varepsilon_k c_k^\dagger c_k 
+  \sum_p \varepsilon_p c_p^\dagger c_p \nonumber \\ 
&& {} \! + (\Gamma 
(\sum_k \gamma_k c_k^\dagger)(\sum_p \gamma_p c_p) + {\rm H.c.} )
 \nonumber \\
&=& \sum_\alpha \varepsilon_\alpha c_\alpha^\dagger c_\alpha
 + \varepsilon_a a_0^\dagger a_0
 + (\sum_\alpha \Gamma_\alpha a_0^\dagger c_\alpha + {\rm H.c.})  \nonumber \\
&& {} \! + \sum_p \varepsilon_p c_p^\dagger c_p
+ (\sum_p \Gamma_p a_0^\dagger c_p + {\rm H.c.}), \nonumber \\
H_{\rm ev} \nonumber &=& 
\hbar \omega (b_0^\dagger b_0 + \frac{1}{2}) 
    + \delta \varepsilon (b_0^\dagger +b_0) a_0^\dagger a_0   \nonumber \\
&& {} \! +\sum_{r} \hbar \omega (b_r^\dagger b_r + \frac{1}{2}) 
+ \sum_{r} \delta \varepsilon (b_r^\dagger +b_r) a_r^\dagger a_r,
\end{eqnarray} 
with $\varepsilon_a = \sum_k |\gamma_k|^2 \varepsilon_k$ and
$\Gamma_p = \Gamma \gamma_p$, 
where  $c_p$  is the annihilation operators for 
the electronic states in the conduction band of the STM tip,
and
$c_k$ is 
the annihilation operator for the electronic states 
in the $\pi$-band for $V_s < 0$ or in the $\pi^*$-band for $V_s > 0$.
The operators of $a_0=\sum_k \gamma^* c_k$ and 
$a_r=\sum_k \exp(-ik \cdot r) \gamma^* c_k$
are the annihilation  operators for the electronic states 
$|a_0 \rangle$ spatially localized
on the dimer beneath the tip and the electronic states 
$|a_r \rangle$ spatially localized
on the dimer at the position $r$ ($r \ne 0$) measured from the dimer beneath the tip, respectively.
The operators of 
$b_0$ and $b_r$ represent the annihilation operators
for the vibrational
state in the rocking mode of the dimer beneath
the tip and of the dimer at the position $r$, respectively.
We assume that the vibrational
states on the dimers are described as the harmonic oscillator.
The energy of the vibrational state $\hbar \omega$ is roughly
estimated as 13meV~\cite{kawai2}. 
The diagonalized term of 
$\sum_k \varepsilon_k c_k^\dagger c_k$
is rewritten by $a_0$ and $c_\alpha$ 
as the first three terms in the third line of eq. (1),
where $c_\alpha$ 
is the annihilation operators for the
electronic states orthogonalized to $|a_0 \rangle$.
The term of 
$\delta \varepsilon$ represents
the electron-vibration coupling.
The vibration is excited or deexcited through 
the electron-vibration coupling by the STM current. 
The essentially same Hamiltonian as eq. (1) has been used in
the previous  study for Si(001) surface~\cite{kawai1} and
other 
surfaces~\cite{other1,other2,other3,other4,other5,other6,other7,other8}.

In the present study, we discuss the transition rates of the vibration 
mainly in the case of $V_s >0$.
The inter-band transition of the electrons
from the conduction band
of the tip to the $\pi^*$-band
can excite coherently the vibrational state of the dimer
through the electron-vibration coupling.
In the coherent transition,
the vibrational state is excited quantumly to the excited state of the higher
quantum number 
through the t-matrix.
The rate of the coherent vibrational excitation 
is proportional to the electronic current
$I_{\rm STM}$~\cite{other4,coh1}. 
The coherent excitation rate $\sigma_{0, 0 \to n}$
in the dimer beneath the tip
from the ground state
of the vibrational number $0$  
to the excited state of $n$ coupled
with the inelastic electrons transition  
is obtained in the lowest order in $|\delta \varepsilon|$
as
\begin{eqnarray}
&&\sigma_{0, 0 \to n}\nonumber \\
&&=2\frac{2\pi}{\hbar} \sum_{\alpha',p'} (1-f(\varepsilon_{\alpha'}-\varepsilon_{\rm F}))f(\varepsilon_{p'}-eV_s-\varepsilon_{\rm F})\nonumber \\
&& \quad \times 
|\langle \alpha',n,0|
H_{\rm ev}(1+G_{\rm e-v}(\varepsilon_p')H_{\rm ev})
|p',0,0 \rangle|^2 \nonumber \\
&& \quad \times 
\delta (\varepsilon_{\alpha'} + n\hbar \omega-\varepsilon_{p'}) 
\nonumber \\
&&= n! \frac{4|\delta \varepsilon |^{2n}}{\pi \hbar}
\int_{E_{\pi^*}}^{eV_s-n\hbar\omega} {\rm d} \varepsilon
\Delta_t(\varepsilon+n\hbar\omega)\Delta_s(\varepsilon) \nonumber \\
&& \quad \times
\prod_{j=0}^n |G_{00}(\varepsilon+j\hbar\omega)|^2 \nonumber \\
&&= n! \frac{4|\delta \varepsilon |^{2n}}{\pi \hbar}
\Delta_t\Delta_s(eV_s-n\hbar\omega-\lambda_n) \nonumber \\
&& \quad \times
\prod_{j=0}^n |\tilde{G}_{00}(eV_s-\lambda_n-j\hbar\omega)|^2 
\nonumber \\
&& \quad \times (eV_s-n\hbar\omega-E_{\pi^*}), \nonumber \\
&&G_{\rm e-v}(\varepsilon) = (\varepsilon+i0^+ -(H_{\rm e}+H_{\rm ev}))^{-1},
\nonumber \\
&&G_{00}(\varepsilon)= 
\langle a_0 |(\varepsilon+i0^+ -H_{\rm e})^{-1} | a_0 \rangle \nonumber \\
&&\quad \quad \quad = (\varepsilon - \varepsilon_a - \Lambda(\varepsilon)+i\Delta(\varepsilon))^{-1},
\end{eqnarray}
where $f(\varepsilon)=(\exp(\beta\varepsilon)+1)^{-1}$,
$\tilde{G}_{00}(\varepsilon) =
(\varepsilon - \varepsilon_a + i\Delta_s(\varepsilon))^{-1}$
and
$\varepsilon_{\rm F}$ is the Fermi level of the electronic system
in the surface.
$G_{00}$ and $G_{\rm e-v}$ are the Green functions for the electronic system
and for the whole system, respectively. 
$\Lambda(\varepsilon)$ is 
$\pi^{-1} 
P\int_{-\infty}^{\infty} \! {\rm d} \varepsilon' 
\Delta(\varepsilon')/(\varepsilon-\varepsilon')$
where $P$ denotes the Cauchy principal value. 
$\Delta(\varepsilon)=\Delta_s(\varepsilon)+\Delta_t(\varepsilon)$ 
is the width of
projected density of states for $| a_0 \rangle$. 
$\Delta_s(\varepsilon)=\pi \sum_\alpha |\Gamma_\alpha|^2 
\delta (\varepsilon - \varepsilon_\alpha)$ 
and $\Delta_t(\varepsilon)=\pi \sum_p |\Gamma_p|^2 
\delta (\varepsilon - \varepsilon_p)$ 
are the components of $\Delta(\varepsilon)$ in the surface
and in the tip, respectively.
$E_{\pi^*}$ is the energy level
of the bottom of the $\pi^*$-band, and  $D_{\pi^*}$ is the width
of the $\pi^*$-band. 
Within the $\pi^*$-band, 
$\Delta(\varepsilon)$ is approximated 
as $\Delta(\varepsilon) \approx \Delta_s(\varepsilon)$,
because $|\Gamma_\alpha|$ is to be 
much larger than $|\Gamma_p|$ in the STM
experiments.  
$|\alpha' \rangle$ and
$|p' \rangle$ are the stationary states
of $H_{\rm e}$ connected to 
$|\alpha \rangle$ and
$|p \rangle$, respectively.
Namely, the stationary states are given as
$|\alpha' \rangle = |\alpha \rangle + 
(\varepsilon_\alpha+i0^+ - H_{\rm e})^{-1} 
\Gamma_\alpha | a_0 \rangle $ 
and 
$|p' \rangle = |p \rangle + (\varepsilon_p+i0^+ - H_{\rm e})^{-1} 
\Gamma_p | a_0 \rangle $. 
The broadness of the Fermi distribution around
$\varepsilon_{\rm F}$ in the $\pi^*$ band
and around  $\varepsilon_{\rm F}+V_s$ in the tip
is assumed to be negligible in the evaluation
of the integration in eq. (2).
The evaluation point for the integration in eq. (2) 
are defined at $\varepsilon=eV_s-n\hbar\omega-\lambda_n$.
$\Delta_t(\varepsilon)$ is assumed  to take constant value of
$\Delta_t$ in the evaluations in eq. (2).
Because the height of the energy barrier for the flip-flop motion
of the dimer in the c(4$\times$2) ordered structure and the height
for the type-P defect are estimated from the model potential
of Ge(001)~\cite{kawai2} as about $4.2\times10^2$meV and
$3.1\times10^2$meV, respectively,
$n$ is estimated to be of the order of 
20 $\sim$ 30 for the transformations. 

The surface localized states of the $\pi$-band and the $\pi^*$-band
are formed mainly through the hybridization of the dangling bonds on the
dimers. The first-principles calculations (FPC)~\cite{rohlfing1}
and scanning tunneling
spectroscopy (STS)~\cite{komori1,komori2} 
show that the $\pi$-band and the $\pi^*$-band 
have the two-dimensional character and the quasi-one-dimensional
character, respectively, and
the band widths of both bands are
obtained to be about 1 eV. 
In quasi-one-dimensional bands, the density of states 
has large values near the band edges, and 
becomes small around the central region of the band.
In the present study, we assume 
$\varepsilon_a$ of $\pi^*$-band to be the center of the band in energy:
$\varepsilon_a = E_{\pi^*} + D_{\pi^*}/2$. 
Corresponding to the above mentioned character of the
quasi-one-dimensional band, 
the energy dependence of $\Delta_s (\varepsilon)$ is simply modeled
as
\begin{eqnarray}
\Delta_s(\varepsilon) &=& \left\{
\begin{array}{ll}
\Delta_{\rm S}, & (|\varepsilon-\varepsilon_a| < \phi) \\
\Delta_{\rm L}, & (\phi \le |\varepsilon-\varepsilon_a| < \frac{1}{2}D_{\pi^*}) \\
0,              & {\rm otherwise}
\end{array}
\right. .
\end{eqnarray} 
We know by the STS results~\cite{komori1,komori2} 
that $\Delta_{\rm L}/\Delta_{\rm S}$ and
$\phi$ are of the order of 2 and 50meV, respectively.
Because the value of $\Delta_{\rm L}$ is roughly estimated about 500meV
from the width of the $\pi^*$-band,
$\phi$ is expected to be much smaller than $\Delta_{\rm S}$.

In this model, $\sigma_{0, 0 \to n}$
are obtained from eq. (2) for $E_{\pi^*} < eV_s-n\hbar\omega$ as
\begin{eqnarray}
\sigma_{0, 0 \to n} &=& 
n ! \frac{4|\delta \varepsilon|^{2n}}{\pi \hbar} \Delta_t \Delta_{\rm L}
\prod_{j=0}^{\ell_1-1} L_j(\Delta_{\rm L})
\nonumber \\
&& {} \times 
\prod_{j  =\max(0,\ell_1)}^{\ell_2-1}  \! \! \! \! L_j(\Delta_{\rm S})
\prod_{j  =\max(0,\ell_2)}^{n} \! \! \! \! L_j(\Delta_{\rm L})
\nonumber \\
&& {} \times  (eV_s-n\hbar\omega-E_{\pi^*}),
\end{eqnarray}
where $L_j(x)=((ev_s - j \hbar \omega)^2 + x^2)^{-1}$. 
The integers $\ell_1$ and $\ell_2$ are defined by
$(ev_s-\ell_1\hbar \omega) < \phi < (ev_s - (\ell_1-1) \hbar \omega)$
and 
$(ev_s-\ell_2\hbar \omega < - \phi < (ev_s - (\ell_2-1) \hbar \omega)$,
respectively, $\prod_{k}^{\ell}$ is assigned to 1 for $\ell < 0$,
and $ev_s=eV_s-\lambda_n-\varepsilon_a$.
We take the scope to the case of $ev_s \sim 0$, hereafter.
As mentioned above, we assume from STS results~\cite{komori1,komori2} that 
$\phi \ll \Delta_{\rm s} < \Delta_{\rm L}$. In the condition of 
$ev_s \sim 0$, the voltage dependent terms of $ev_s - j \hbar \omega$ 
appearing in the denominators of the products in eq. (4)
are approximately neglected.
The inelestic rate of  $\sigma_{0, 0 \to n}$
are classified by the relative values of $ev_s$ to $-\phi$ or $\phi$ as
\begin{eqnarray}
\sigma_{0, 0 \to n} &\approx&
\frac{4 n!\Delta_t}{\pi \hbar \Delta_{\rm L}}
\left(\frac{|\delta \varepsilon|}{\Delta_{\rm L}}\right)^{2n} (eV_s-n\hbar\omega-E_{\pi^*})
F(ev_s),  \nonumber \\
F(ev_s) &=& \left\{
\begin{array}{ll}
1 & (ev_s \le -\phi)\\
\left(\Delta_{\rm L}/\Delta_{\rm s}\right)^{2(ev_s+\phi)/\hbar\omega} 
& (-\phi \le ev_s < \phi) \\
\left(\Delta_{\rm L}/\Delta_{\rm s}\right)^{4\phi/\hbar\omega} & (\phi \le ev_s )\\
\end{array}
\right. .
\end{eqnarray}
Eq. (5) shows that $\sigma_{0, 0 \to n}$
increases gradually with the powers of the bias voltage for $ev_s < -\phi$, 
and 
that it starts to increase exponentially with the bias voltage
around $ev_s \approx -\phi$.
At the bias voltages  of $ev_s > \phi$,
$\sigma_{0, 0 \to n}$  increases again with the powers of the bias voltage. 
The enhancement for $\sigma_{0, 0 \to n}$ caused by 
the exponential increase in the range of $-\phi \le ev_s < \phi$
is up to $(\Delta_{\rm L}/\Delta_{\rm S})^{4\phi/\hbar \omega}$.
The enhancement factor 
$(\Delta_{\rm L}/\Delta_{\rm S})^{4\phi/\hbar \omega}$
is expected to
reaches
$2^{15} \approx 3.3 \times 10^4$ from the estimated values.
Though the estimations for 
$\Delta_{\rm L}/\Delta_{\rm S}$ and $\phi$
do not have a high accuracy, 
we know that 
the rate of the local transformation  
increases steeply around $eV_s=\varepsilon_a + \lambda_n$,
and the enhancement reaches to the order of $10^4$.
$\lambda_n$ for  the flip-flop motion
and $\varepsilon_a$
are estimated to be about 0.1eV and 0.6eV, respectively,
from the height of the energy barrier for the flip-flop motion
and the width of the $\pi^*$-band and 
the STS results~\cite{komori1,komori2}.
Using the estimated values of $\lambda_n$ and $\varepsilon_a$,
we get $V_s \approx 0.7$V at which the condition $ev_s=0$ is  satisfied. 
This value of 0.7V for the bias voltage shows good agreement with
the experimental results~\cite{komori1,komori2}
where the transformation to the p(2$\times$2) structure for $V_s \le 0.7$V
is so slow as to be hardly observed, and
that for 0.8V $\le V_s$ is  fast enough to be well observed.

The excitation rate of the vibrartion of 
the dimer at the position $r$
away from the dimer beneath the tip, 
$\sigma_{r,0 \to n}$
is obtained in the lowest order in $|\delta \varepsilon|$
in the essentially same way as $\sigma_{0,0 \to n}$: 
\begin{eqnarray}
&&\sigma_{r, 0 \to n}\nonumber \\
&&=2\frac{2\pi}{\hbar} \sum_{\alpha',p'} (1-f(\varepsilon_{\alpha'}-\varepsilon_{\rm F}))f(\varepsilon_{p'}-eV_s-\varepsilon_{\rm F})\nonumber \\
&&\quad \times 
|\langle \alpha',n,r |
H_{\rm ev}(1+G_{\rm e-v}(\varepsilon_p')H_{\rm ev})
|p',0,0 \rangle|^2 \nonumber \\
&& \quad \times 
\delta (\varepsilon_{\alpha'} + n\hbar \omega-\varepsilon_{p'})
\nonumber \\
&&= n! \frac{4|\delta \varepsilon |^{2n}}{\pi \hbar}
\int_{E_{\pi^*}}^{eV_s-n\hbar\omega} {\rm d} \varepsilon
\Delta_t(\varepsilon+n\hbar\omega)\Delta_s(\varepsilon)\nonumber \\
&& \quad \times |G_{r0}(\varepsilon+n\hbar\omega)|^2
\prod_{j=0}^{n-1} 
|G_{rr}(\varepsilon+j\hbar\omega)|^2,
\end{eqnarray}
where
$G_{rr}(\varepsilon)= 
\langle a_r |
(\varepsilon+i0^+ -H_{\rm e})^{-1} 
| a_r \rangle = G_{00}(\varepsilon)$ and
$G_{r0}(\varepsilon)= 
\langle a_r |
(\varepsilon+i0^+ -H_{\rm e})^{-1} 
| a_0 \rangle$. 
In order to obtain $G_{r0}(\varepsilon)$,
we rewrite 
the Hamiltonian $H_{\rm e}$ by the bases of 
$| a_0 \rangle$, $| a_r \rangle$
and their orthogonalized states $| q \rangle$:
\begin{eqnarray}
H_{\rm e} 
&=& \sum_q \varepsilon_q 
c_q^\dagger c_q 
+ \varepsilon_a a_0^\dagger a_0
 + \varepsilon_a a_r^\dagger a_r
\nonumber \\
&&{} + (\sum_q \Gamma_{0q} a_0^\dagger c_q + {\rm H.c.})  
 + (\sum_q \Gamma_{rq} a_r^\dagger c_q + {\rm H.c.}) \nonumber \\
&&{} + ( W_{r0}a_r^\dagger a_0 + {\rm H.c.}),
\end{eqnarray} 
where the terms concerning with the tip are omitted.
The transfer integrals of $\Gamma_{rq}$ and $\Gamma_{0q}$  
in eq. (7) have the magnitudes of same order as  $\Gamma_\alpha$ in
eq. (1) for large $r$.
Using the rewritten form of $H_{\rm e}$ in eq. (7),
we obtain easily $G_{r0}(\varepsilon)$ as
$G_{r0}(\varepsilon)=(\varepsilon-\varepsilon_a+i0^+ - C_{00})^{-1}
(C_{r0}+W_{r0})G_{00}$,
where $C_{00}=\sum_q |\Gamma_{0q}|^2(\varepsilon-\varepsilon_q+i0^+)^{-1}$ 
and
$C_{r0}=
\sum_q \Gamma_{rq}\Gamma_{0q}^*(\varepsilon-\varepsilon_q+i0^+)^{-1}$.
The integration in eq. (6) is approximately evaluated
at $eV_s-n\hbar \omega -\lambda_n$ same way as for $\sigma_{0, 0 \to n}$.
The excitation rate $\sigma_{r, 0 \to n}$ is approximately obtained as
\begin{eqnarray}
\sigma_{r, 0 \to n} &\approx& \
|(-i\Delta_{r0}+W_{r0})|^2 \Delta_{\rm 00}^{-2} \sigma_{0, 0 \to n},
\end{eqnarray} 
where $\Delta_{00}= \pi \sum_q |\Gamma_{0q}|^2 
\delta (ev_s+\varepsilon_a-\varepsilon_q)$ and
$\Delta_{r0}= \pi \sum_q \Gamma_{rq}\Gamma_{0q}^*
\delta (ev_s+\varepsilon_a-\varepsilon_q)$.
The absolute value of $W_{r0}$ almost vanishes out
within the distance of a few dimers. 
For the one-dimensional band, 
both the absolute values of $\Delta_{r0}$ and $W_{r0}$ 
vanish  on the dimer rows  
other than the row including the dimer beneath the tip.
On the dimer row including the dimer beneath the tip,
the absolute value  of $\Delta_{r0}$ does not decay 
with the absolute value of $r$
for the one-dimensional band, 
because 
only a few discrete points of $q$
contribute to the summation of $q$ for $\Delta_{r0}$ for the one-dimensional band. 
We know that
because of the quasi-one-dimensional charactoer of the $\pi^*$-band,
the transformation  to the p(2$\times$2) structure for $V_s > 0$
is restricted  almost into the dimer row including the dimer beneath the
tip and the rate of the transformation decays much slowly with the distance
from the dimer beneath the tip.
The transformed region into the p(2$\times$2) structure along the dimer row
is expected to
extend to be of 
the order of the mean free path of the electron in the $\pi^*$-band. 

For the two-dimensional band, the excitation rate 
$\sigma_{r,0 \to n}$ is also obtained in the same form as in eq. (8).
In spite of the fast decay of the excitation rate on $r$,
$W_{r0}$ does not vanish on both the direction of parallel and 
perpendicular to the dimer row in the two demensional band.
The absolute value of $W_{r0}$ is  effective
only within the distance of a few dimers. 
For the two-demensional band, 
$\Delta_{r0}$ shows the quasi-isotropic power-law decay on $r$,  
because in the two-dimensional band, the  continuous points of $q$ on a
line in the $q$-space
contribute to the summation of $q$ for
$\Delta_{r0}$,
and 
the summation of $q$
is replaced by integral on the line in the $q$-space.
These results for the position dependence of the rate 
for the two demensional band reproduce well the experiment
results~\cite{komori1,komori2} of the local transformation   to  
the c(4$\times$2) for $V_s < 0$,
where the transformed region is restricted within the range of a few nm
away from the dimer beneath the tip 
and 
the transformed region has the quasi-isotropic shape.

In conclusion, we have investigated theoretically 
the coherent excitation rate of the dimer vibration,
using the Hamiltonian consisting of the terms
of the electron system and the electron-vibration coupling terms.
We show that
the transformations of the local structures by
STM~\cite{komori1,komori2}
is driven by the dimer vibration excited coherently by the STM current. 
The sample bias voltage 
above which the transformation to the p(2$\times$2) structure
is able to be observed, is semiquantitatively reproduced 
by the quasi-one-dimensional character of the $\pi^*$-band.
For the one-dimensional band,
the excitation rate is shown to have a term not decaying 
on the distance from the
STM tip.  
The contrastive diffrence of the shapes of the transformed region depending
on the sign of the bias voltage observed 
by the experiments~\cite{komori1,komori2}
is explained by the dimensionality of  the $\pi$-band and 
the $\pi^*$-band.

\end{document}